\documentclass[epj]{svjour}

\usepackage{graphicx}
\usepackage{fancyhdr}
\def\makeheadbox{{
 }}
\def\mytitle{My title} 
\def\myauthors{My name}  
\def\mytype{My type of session}
\def\mysession{My session}

\def\mytitle{Leptogenesis in $SO(10)$ models with a left-right symmetric
seesaw mechanism} 
\def\myauthors{St\'ephane Lavignac}    
\def\mytype{Contributed Talk}    
\def\mysession{Flavor Physics}

\begin{document}
\title{Leptogenesis in $SO(10)$ models with a left-right symmetric
seesaw mechanism
}
\author{A. Abada\inst{1}, P. Hosteins\inst{2},
F.-X. Josse-Michaux\inst{1} and S. Lavignac\inst{3}
\thanks{\emph{Email:} Stephane.Lavignac@cea.fr} %
}                     
%
%
\institute{Laboratoire de Physique Th\'eorique,
Universit\'e de Paris-Sud, B\^atiment 210,
F-91405 Orsay Cedex, France
\and Department of Physics,
University of Patras,
GR-26500 Patras, Greece
\and Service de Physique Th\'eorique,
Orme des Merisiers, CEA-Saclay,
F-91191 Gif-sur-Yvette Cedex, France}
%
\date{}
\abstract{
We study leptogenesis in supersymmetric $SO(10)$ models with a
left-right symmetric seesaw mechanism, including flavour effects
and the contribution of the next-to-lightest right-handed neutrino.
Assuming $M_D = M_u$ and hierarchical light neutrino masses,
we find that successful leptogenesis is possible
for 4 out of the 8 right-handed neutrino mass spectra that are
compatible with the observed neutrino data. An accurate description
of charged fermion masses appears to be an important ingredient
in the analysis.
} 
\maketitle
%


\section{Introduction}
\label{sec:intro}

Testing the seesaw mechanism~\cite{seesaw} is almost certainly
an hopeless goal, except for specific low-energy realizations. The main
reasons we have to believe in it are its elegance and the fact that it fits
so nicely into $SO(10)$ unification.
This motivates us to investigate its observable implications,
such as leptogenesis~\cite{FY86} and, in supersymmetric
theories, lepton flavour violation.

So far most studies of leptogenesis have been done in the framework
of the type I (heavy right-handed neutrino exchange) seesaw mechanism,
or assumed dominance of either the type I or the type II (heavy scalar
$SU(2)_L$ triplet exchange) seesaw mechanism. It is interesting,
though, to investigate whether the generic situation where both
contributions are comparable in size can lead to qualitatively
different results. A further motivation to do so comes from
the well-known fact that successful leptogenesis is difficult
to achieve in $SO(10)$ models with a type I seesaw mechanism,
which generally\footnote{This might not be the case in models
where the relation $M_D = M_u$ receives large corrections from
Yukawa couplings involving a $\bf \overline{126}$ or $\bf 120$
Higgs representation, or from non-renormalizable interactions.}
present a very hierarchical right-handed neutrino mass spectrum,
with $M_1$ lying below the Davidson-Ibarra bound~\cite{DI02}.

In this talk, we present results on leptogenesis in
$SO(10)$ models with a left-right symmetric seesaw mechanism.
Details can be found in Refs.~\cite{HLS,AHJL} (for related work,
see Refs.~\cite{ABHKO06,Hallgren07}).


\section{Right-handed neutrino spectra in the left-right symmetric
seesaw mechanism}
\label{sec:reconstruction}


\subsection{The left-right symmetric seesaw mechanism}

In left-right symmetric extensions of the Standard Mo\-del,
the light neutrino mass matrix is often given by the following
formula~\cite{seesawII}:
\begin{equation}
  M_\nu\ =\ f v_L - \frac{v^2}{v_R} Y^T_\nu f^{-1} Y_\nu\ .
\label{eq:LRseesaw}
\end{equation}
In Eq.~(\ref{eq:LRseesaw}), $v_R$ is the scale of $B-L$ breaking,
$v$ is the electroweak scale, and
$v_L \sim v^2 v_R / M^2_{\Delta_L}$
is the vev of the heavy $SU(2)_L$ triplet.
A discrete left-right symmetry ensures that
a single symmetric matrix $f$ determines both the couplings
of the $SU(2)_L$ triplet to lepton doublets, to which the type II contribution
(first term) is proportional, and the right-handed neutrino mass matrix
$M_R = f v_R$, which enters the type I contribution (second term).
The discrete symmetry also constrains the Dirac coupling matrix
$Y_\nu$ to be symmetric.

\begin{figure*}
\begin{center}
\includegraphics[width=0.3\textwidth]{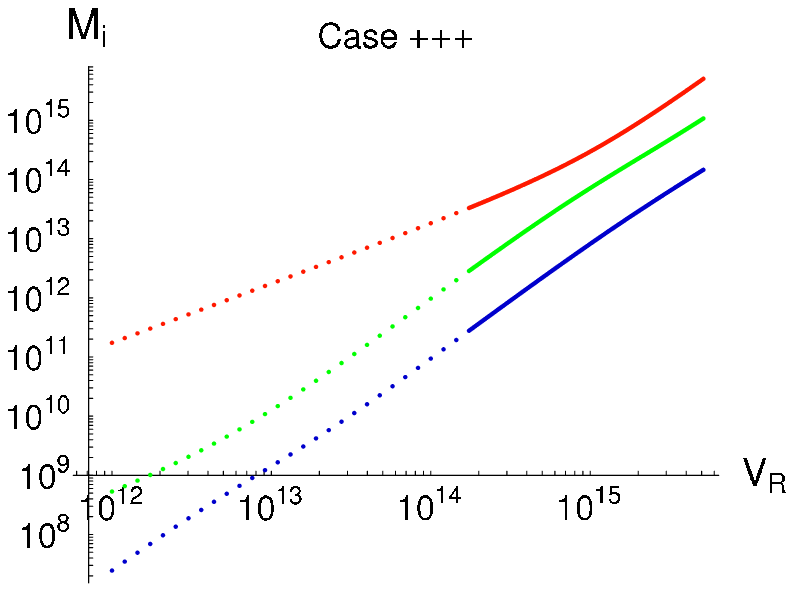}
\hskip .5cm
\includegraphics[width=0.3\textwidth]{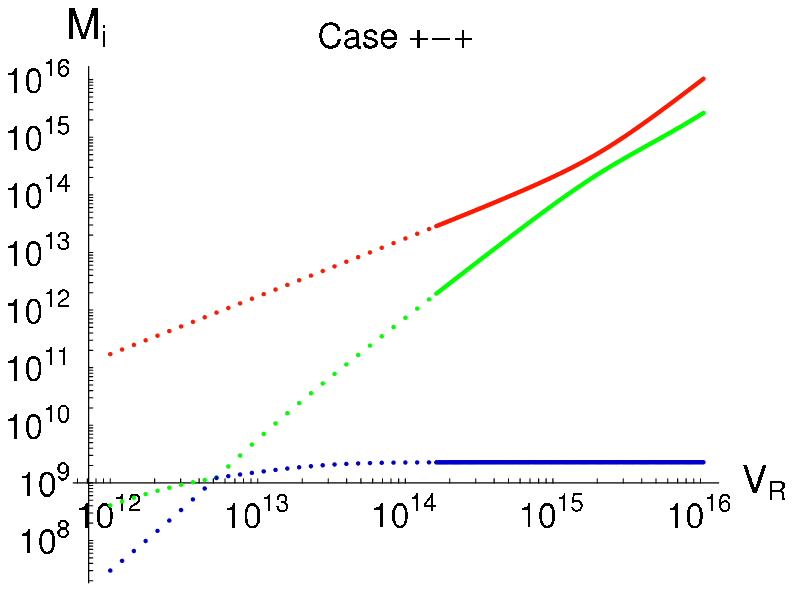}
\hskip .5cm
\includegraphics[width=0.3\textwidth]{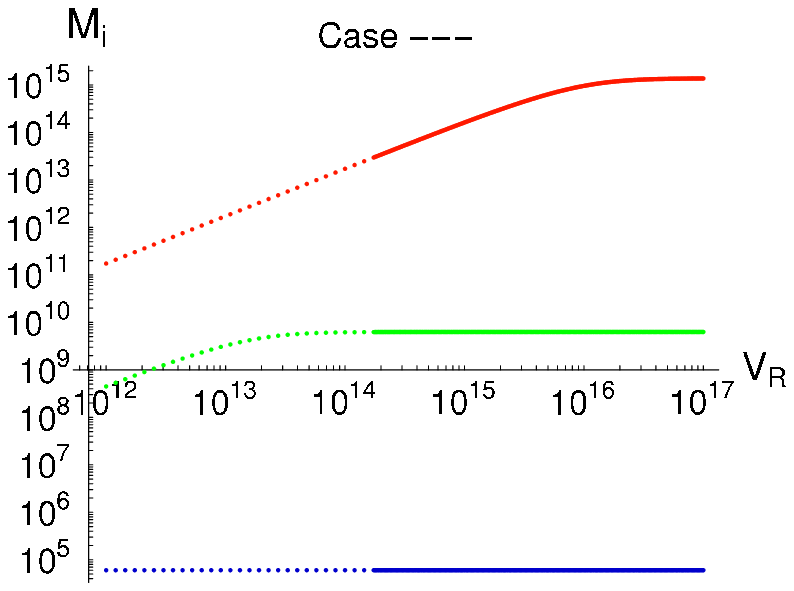}
\end{center}
\vskip -.1cm
\caption{Right-handed neutrino masses as a function of $v_R$ (in GeV)
for solutions $(+,+,+)$ {\it (left)}, $(+,-,+)$ {\it (middle)} and
$(-,-,-)$ {\it (right panel)}. Inputs: hierarchical light neutrino masses
with $m_1 = 10^{-3}$ eV, $\sin^2 \theta_{13} = 0.009$, $\beta / \alpha = 0.1$
and no CP violation beyond the CKM phase. The range of variation of $v_R$
is restricted from above by the requirement that $f_3 \leq 1$. Dotted lines
indicate a fine-tuning greater than 10\% in the $(3,3)$ entry of the light
neutrino mass matrix.}
\label{fig:spectra}       
\end{figure*}

In order to study leptogenesis, the knowledge of the masses and couplings
of the right-handed neutrinos and of the $SU(2)_L$ triplet is needed.
Therefore, in a theory which predicts the Dirac matrix $Y_\nu$,
one must solve Eq.~(\ref{eq:LRseesaw}) for the $f_{ij}$
couplings, assuming a given pattern for the light neutrino masses
and mixings. In Ref.~\cite{AF05}, it was shown that this ``reconstruction''
problem has exactly $2^n$ solutions for $n$ families,
and explicit expressions for the $f_{ij}$'s were provided up to
$n = 3$. Here we use the alternative reconstruction
procedure proposed in Ref.~\cite{HLS}.


\subsection{Reconstruction procedure}

In order to solve Eq.~(\ref{eq:LRseesaw}), we first rewrite it as
\begin{equation}
  Z\ =\ \alpha X - \beta X^{-1}\, ,
\label{eq:master_equation}
\end{equation}
with $\alpha \equiv v_L$, $\beta \equiv v^2 / v_R$ and
\begin{equation}
  Z\ \equiv\ N_\nu^{-1} M_\nu (N_\nu^{-1})^T\, , \quad
  X\ \equiv\ N_\nu^{-1} f (N_\nu^{-1})^T\, ,
\label{eq:def_Z_X}  
\end{equation}
where $N_\nu$ is a matrix such that $Y_\nu = N_\nu N_\nu^T$,
and $Y_\nu$ is assumed to be invertible.
Being complex and symmetric, $Z$ can be diagonalized by a complex
orthogonal matrix if its eigenvalues (i.e. the roots of the
characteristic polynomial $\det (Z - z \mathbf{1}) = 0$) are all distinct:
\begin{equation}
 Z\ =\ O_Z \mbox{Diag}\, (z_1, z_2, z_3) O^T_Z\ ,  \qquad  O_Z O^T_Z\ =\ \mathbf{1}\ .
\label{eq:diag_Z} 
\end{equation}
Then, upon an $O_Z$ transformation, Eq. (\ref{eq:master_equation})
reduces to 3 independent quadratic equations for the eigenvalues of $X$,
$z_i = \alpha x_i - \beta x^{-1}_i$.  For a given choice of ($x_1$, $x_2$, $x_3$),
the solution of Eq.~(\ref{eq:LRseesaw}) is given by:
\begin{equation}
  f\ =\ N_\nu\, O_Z\, \mbox{Diag}\, (x_1, x_2, x_3)\, O^T_Z\, N_\nu^T\ .
\label{eq:f}
\end{equation}
The right-handed neutrino masses $M_i = f_i v_R$ are obtained
by diagonalizing $f$ with a unitary matrix $U_f$,
and the couplings of the right-handed neutrino mass eigenstates
are given by $Y \equiv U^\dagger_f Y_\nu$.

Since each equation $z_i = \alpha x_i - \beta x^{-1}_i$ has
two solutions $x^-_i$ and $x^+_i$, there are 8 different
solutions for the matrix $f$, which we label in the following way:
$(+,+,+)$ refers to the solution $(x^+_1, x^+_2, x^+_3)$, $(+,+,-)$
to the solution $(x^+_1, x^+_2, x^-_3)$, and so on.
It is convenient to define $x^-_i$ and $x^+_i$ such that, in the
$4 \alpha \beta \ll |z_i|^2$ limit:
\begin{equation}
  x^-_i\ \simeq\ - \frac{\beta}{z_i}\ ,  \qquad  x^+_i\ \simeq\ \frac{z_i}{\alpha}\ .
\label{eq:x_pm_limit}
\end{equation}
With this definition, the large $v_R$ limit ($4 \alpha \beta \ll |z_1|^2$)
of solutions $(-,-,-)$ and $(+,+,+)$ corresponds to the ``pure'' type I
and type II cases, respectively:
\begin{eqnarray}
  f^{(-,-,-)} & \stackrel{4 \alpha \beta \ll |z_1|^2}{\longrightarrow} &
  -\, \frac{v^2}{v_R}\, Y_\nu M^{-1}_\nu Y_\nu\ ,  \\
  f^{(+,+,+)} & \stackrel{4 \alpha \beta \ll |z_1|^2}{\longrightarrow} &
  \frac{M_\nu}{v_L}\ .
\end{eqnarray}
The remaining 6 solutions correspond to mixed cases where
the light neutrino mass matrix receives significant contributions
from both types of seesaw mechanisms.
In the opposite, small $v_R$ limit ($|z_3|^2 \ll 4 \alpha \beta$), one has
$x^\pm_i\ \simeq\ \pm\, \mbox{sign} (\mbox{Re} (z_i)) \sqrt{\beta / \alpha}$,
which indicates a partial cancellation between the type I and type II
contributions to light neutrino masses.

\begin{figure*}
\begin{center}
\includegraphics[width=0.3\textwidth]{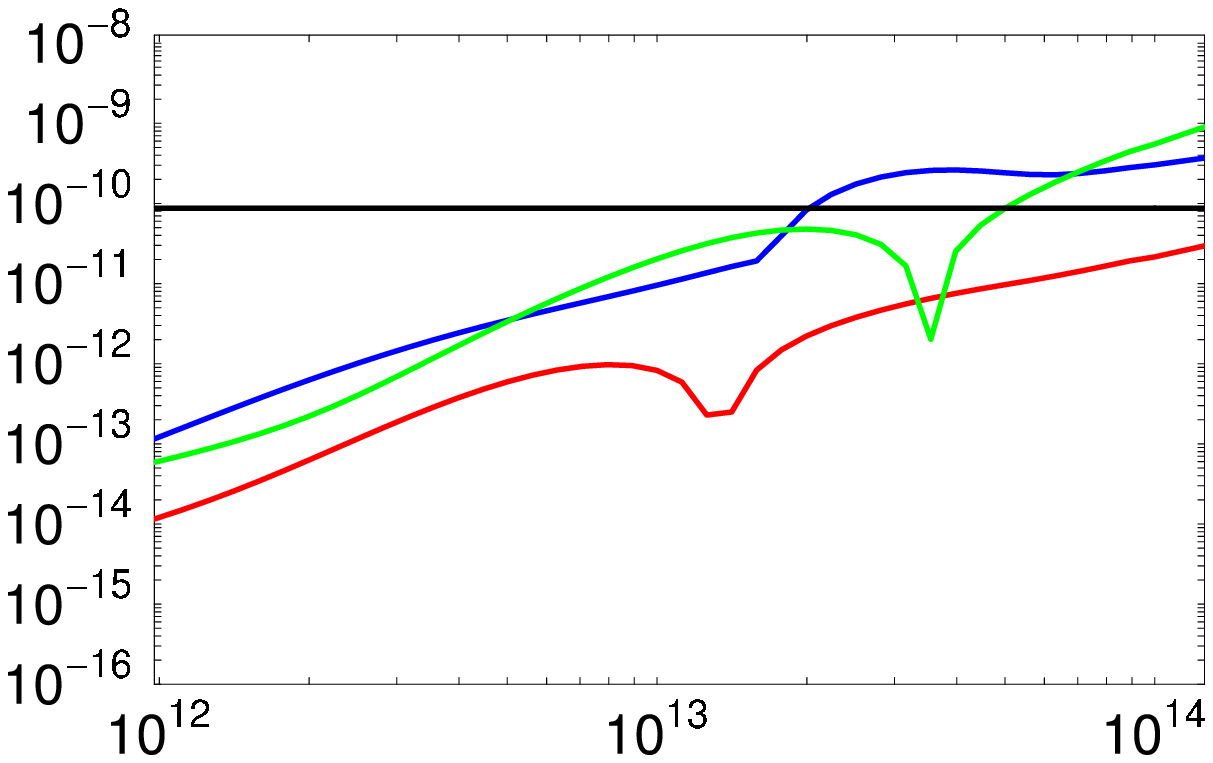}
\hskip .5cm
\includegraphics[width=0.3\textwidth]{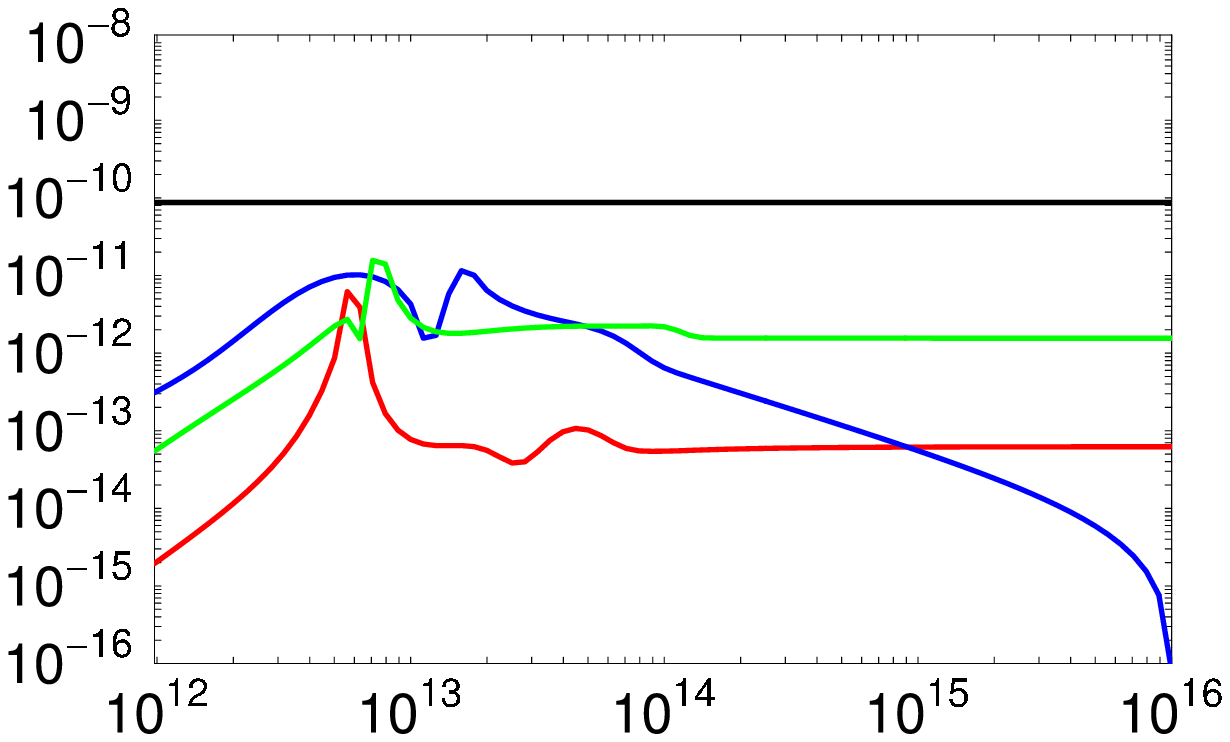}
\hskip .5cm
\includegraphics[width=0.3\textwidth]{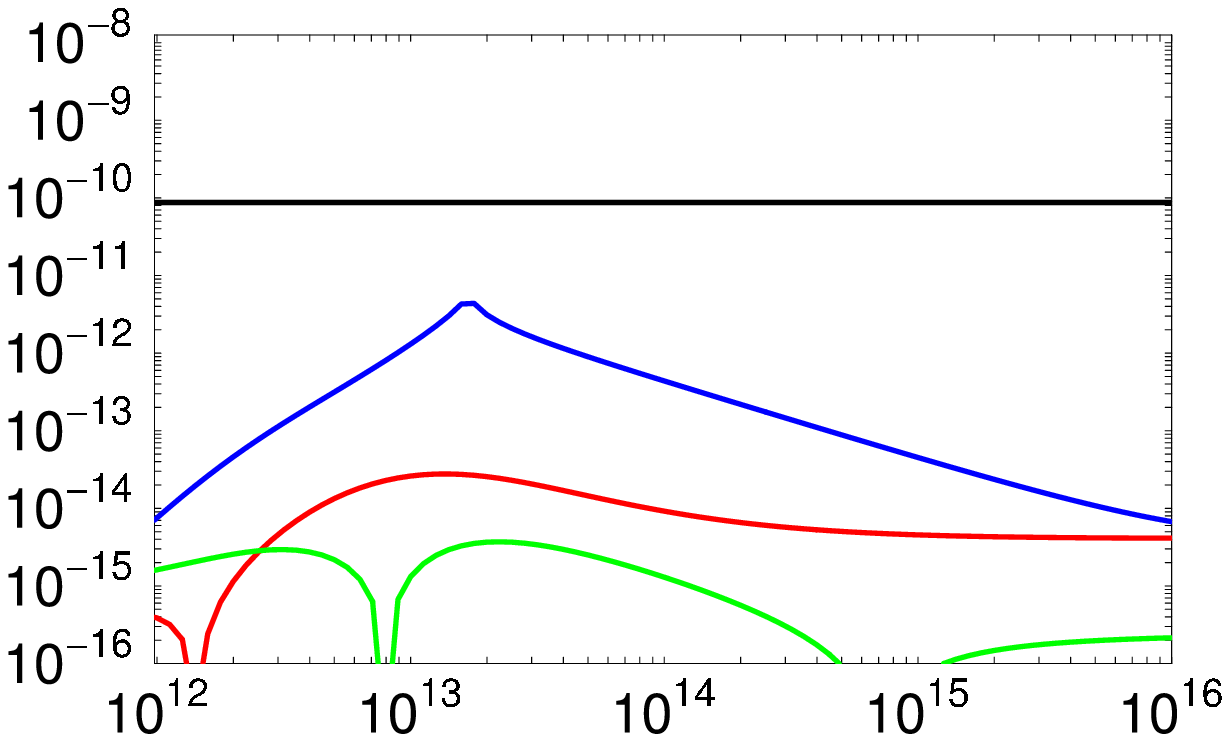}
\end{center}
\vskip -.1cm
\caption{$Y_B$ as a function of $v_R$ (in GeV) for solutions
$(+,+,+)$ {\it (left)}, $(+,-,+)$ {\it (middle)} and $(-,-,-)$
{\it (right panel)}. Inputs: hierarchical light neutrino mass spectrum
with $m_1 = 10^{-3}$ eV, $\sin^2 \theta_{13} = 0.009$ and $\delta_{PMNS} = 0$;
$\beta / \alpha = 0.1$; three different choices for the Majorana and
high-energy phases {\it (blue: $\Phi^u_2 = \pi/4$; green: $\Phi^\nu_2 = \pi/4$;
red: no CP violation beyond the CKM phase)}; vanishing initial abundance
for $N_1$ and $N_2$.}
\label{fig:YB}       
\end{figure*}


\subsection{Application to $SO(10)$ models}

Let us now apply the reconstruction procedure to
supersymmetric $SO(10)$ models
with two $\bf 10$s, a $\bf 54$ and a $\bf \overline{126}$
representations in the Higgs sector. The two $\bf 10$s
generate the charged fermion masses, leading to the
well-known relations:
\begin{equation}
  M_u\ =\ M_D\ (\equiv Y_\nu v_u)\ , \qquad M_d\ =\ M_e\ .
\label{eq:M_10}
\end{equation}
The $\bf 54$ and the $\bf \overline{126}$ contain the
$SU(2)_L \times SU(2)_R \times U(1)_{B-L}$ representations
needed for the left-right symmetric seesaw mechanism.
In particular, the $SU(2)_L$ triplet as well as the $SU(2)_R$ triplet
whose vev $v_R$ breaks $B-L$
are components of the $\bf \overline{126}$.
The equality $f_L = f_R$ and the symmetry of $Y_\nu$ are
ensured by $SO(10)$ gauge symmetry.

Then, for a given choice of the light neutrino mass parameters
and of the high energy phases contained in $M_u$, the matrix $Z$
is known\footnote{The implicit additional inputs are $\tan \beta$
(we choose $\tan \beta = 10$) and the values of the up quark masses
and of the CKM matrix at the seesaw scale.} and $f$
can be reconstructed as a function of the $B-L$ breaking scale $v_R$
and of $\beta / \alpha$. Perturbativity of the
$f_{ij}$ couplings constrains $\beta / \alpha \leq {\cal O} (1)$
and restricts the range of $v_R$ from above.
In Fig.~\ref{fig:spectra}, we show the right-handed neutrino mass
spectrum of three representative solutions as a function of $v_R$
for a hierarchical light neutrino mass spectrum.
The 4 solutions with $x_3 = x^-_3$ are
characterized by a constant value of the lightest right-handed
neutrino mass, $M_1 \approx 6 \times 10^4$ GeV;
the 2 solutions with $x_3 = x^+_3$
and $x_2 = x^-_2$ by $M_1 \approx 2 \times 10^{9}$ GeV; and
the 2 solutions with $x_3 = x^+_3$ and $x_2 = x^+_2$ by a rising $M_1$.


\section{Implications for leptogenesis}
\label{sec:lepto}

Since $M_{\Delta_L} \sim (\beta / \alpha)\, v_R$ and $M_1 \ll v_R$
in all solutions, one can safely assume that the $SU(2)_L$ triplet
is heavier than the lightest right-handed neutrino.
Then the dominant contribution to leptogenesis comes from
out-of-equilibrium decays of $N_1$
(in some cases to be discussed below, the next-to-lightest neutrino
$N_2$ will also be relevant). The CP asymmetry in $N_1$ decays,
$\epsilon_{N_1} \equiv 
\left[ \Gamma (N_1 \rightarrow l H)
- \Gamma (N_1 \rightarrow \bar l H^\star) \right]  / 
\left[ \Gamma (N_1 \rightarrow l H) \right. $
$\left. + \Gamma (N_1 \rightarrow \bar l H ^\star) \right]$, receives
two contributions: the standard type I contribution
$\epsilon^I_{N_1}$~\cite{FY86,epsilonI}, and an additional
contribution $\epsilon^{II}_{N_1}$ from a vertex diagram containing
a virtual triplet~\cite{epsilonII,HS03}:
\begin{equation}
  \epsilon^I_{N_1}\ =\ \frac{1}{8 \pi}\ \sum_k\
    \frac{\mbox{Im} \left[ (Y Y^\dagger)_{1k} \right]^2}{(Y Y^\dagger)_{11}}\
    f(x_k)\ ,
\label{eq:epsilon_I}
\end{equation}
\begin{equation}
  \epsilon^{II}_{N_1}\ =\ \frac{3}{8 \pi}\ \sum_{k,l}\
    \frac{\mbox{Im} \left[ Y_{1k} Y_{1l} f^\star_{kl} v^\star_L\right]}
    {(Y Y^\dagger)_{11}}\ \frac{M_1}{v^2_u}\ g(x_\Delta)\ ,
\label{eq:epsilon_II}
\end{equation}
where $f(x) = - \sqrt{x} \left[\, 2 / (x-1) + \ln (1+1/x)\, \right]$,
$g(x) = x \ln (1+1/x)$, $x_k \equiv M^2_k / M^2_1$,
$x_\Delta = M^2_{\Delta_L} /M^2_1$,
and $Y \equiv U^\dagger_f Y_\nu$.
The final baryon asymmetry is given by:
\begin{equation}
  Y_B\ \equiv\ \frac{n_B}{s}\ =\ - 1.48 \times 10^{-3}\, \eta\, \epsilon_{N_1}\ ,
\label{eq:n_B_s}
\end{equation}
where $\eta$ is an efficiency factor to be determined by integrating
the Boltzmann equations. For leptogenesis to be successful,
Eq.~(\ref{eq:n_B_s}) should reproduce the observed baryon-to-entropy
ratio $Y^{obs.}_B = (8.7 \pm 0.3) \times 10^{-11}$~\cite{WMAP}.

The behaviour of the different solutions can be anticipated
from the observation of the mass spectra in Fig.~\ref{fig:spectra}
~\cite{HLS}. Indeed,
successful leptogenesis requires $|\epsilon_{N_1}| \geq {\cal O} (10^{-7})$,
while for $M_1 \ll M_2, M_{\Delta_L}$ Eqs.~(\ref{eq:epsilon_I})
and~(\ref{eq:epsilon_II}) yield the upper bound~\cite{HS03}:
\begin{equation}
  |\epsilon_{N_1}|\ \leq\ 2 \times 10^{-7} \left( \frac{M_1}
    {10^9\, \mbox{GeV}} \right) \left( \frac{m_{max}}{0.05\, \mbox{eV}} \right) .
\label{eq:epsilon_max}  
\end{equation}
Thus, the 4 solutions with $x_3 = x^-_3$ will fail to generate
the observed baryon asymmetry from $N_1$ decays, a conclusion
that generalizes a well-known fact in the type I case.
However, $N_2$ decays can do the job if they generate
a large asymmetry in a lepton flavour that is only mildly washed out
by $N_1$ decays and inverse decays~\cite{Vives05}. The 2 solutions
with $x_3 = x^+_3$ and $x_2 = x^+_2$ have a rising $M_1$
and should be able to reproduce the observed asymmetry,
as in the pure type II case. Finally,
the situation is less conclusive for the 2 solutions with
$x_3 = x^+_3$ and $x_2 = x^-_2$, for which flavour effects
and the contribution of $N_2$ could be decisive.

It is clear from the above discussion that a careful study of leptogenesis
requires the inclusion of the next-to-lightest right-handed neutrino
and of flavour effects~\cite{BCST00}. As is well known in the type I case,
flavour effects can significantly affect the final baryon asymmetry
if there is a hierarchy between the washout parameters for different
lepton flavours~\cite{flavour}. We performed such an analysis
in Ref.~\cite{AHJL}, and present our results here. Fig.~\ref{fig:YB}
shows the final baryon asymmetry $Y_B$ as a function of $v_R$
for solutions $(+,+,+)$, $(+,-,+)$ and $(-,-,-)$.
Not surprisingly, the $(+,+,+)$ solution leads to successful
leptogenesis; however there is a tension with the upper bound
on the reheating temperature from gravitino overproduction~\cite{gravitino}
above $v_R \approx 3 \times 10^{13}$ GeV, where
$M_1 > 10^{10}$ GeV. By contrast, the solutions $(+,-,+)$
and $(-,-,-)$ fail to reproduce the observed baryon
asymmetry\footnote{In Ref.~\cite{ABHKO06}, a different conclusion
has been obtained for the solution $(+,-,+)$ in the case of an inverted
light neutrino mass hierarchy.} .
In the $(-,-,-)$ case, flavour effects prevent an exponential
washout of the $B-L$ asymmetry generated in $N_2$ decays
($N_1$ decays alone would give $Y_B \sim (10^{-17}
- 10^{-15})$), but this is not sufficient for ``$N_2$ leptogenesis''
to work.

\begin{figure*}
\begin{center}
\includegraphics[width=0.3\textwidth]{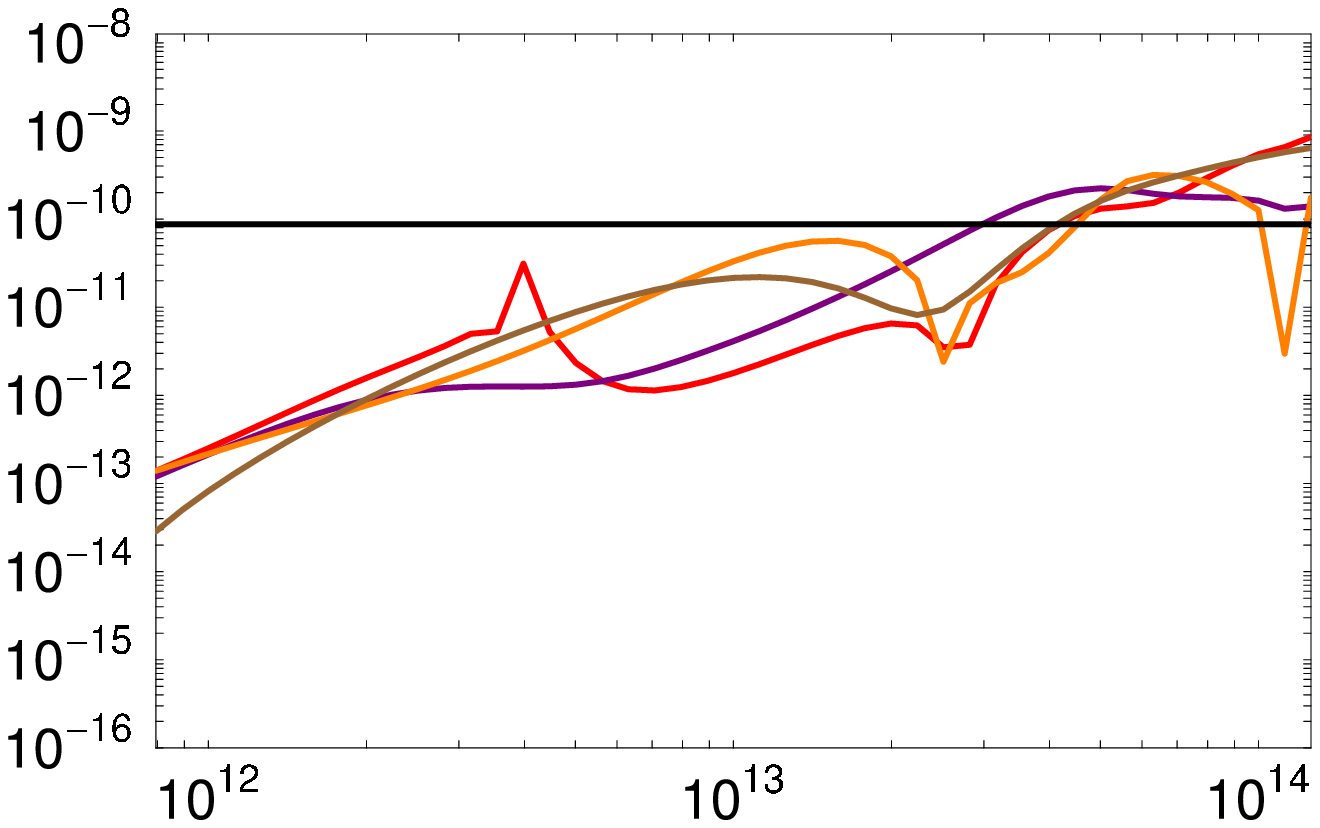}
\hskip .5cm
\includegraphics[width=0.3\textwidth]{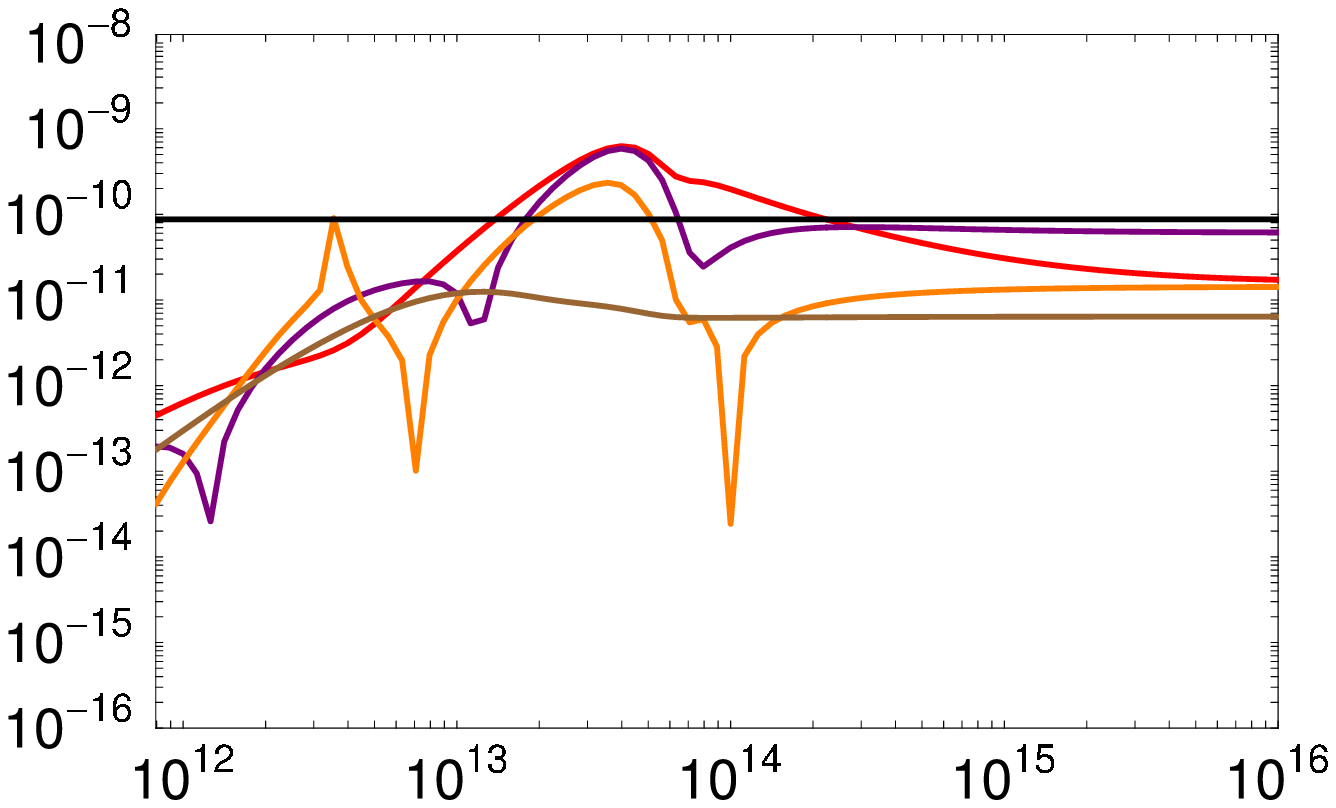}
\hskip .5cm
\includegraphics[width=0.3\textwidth]{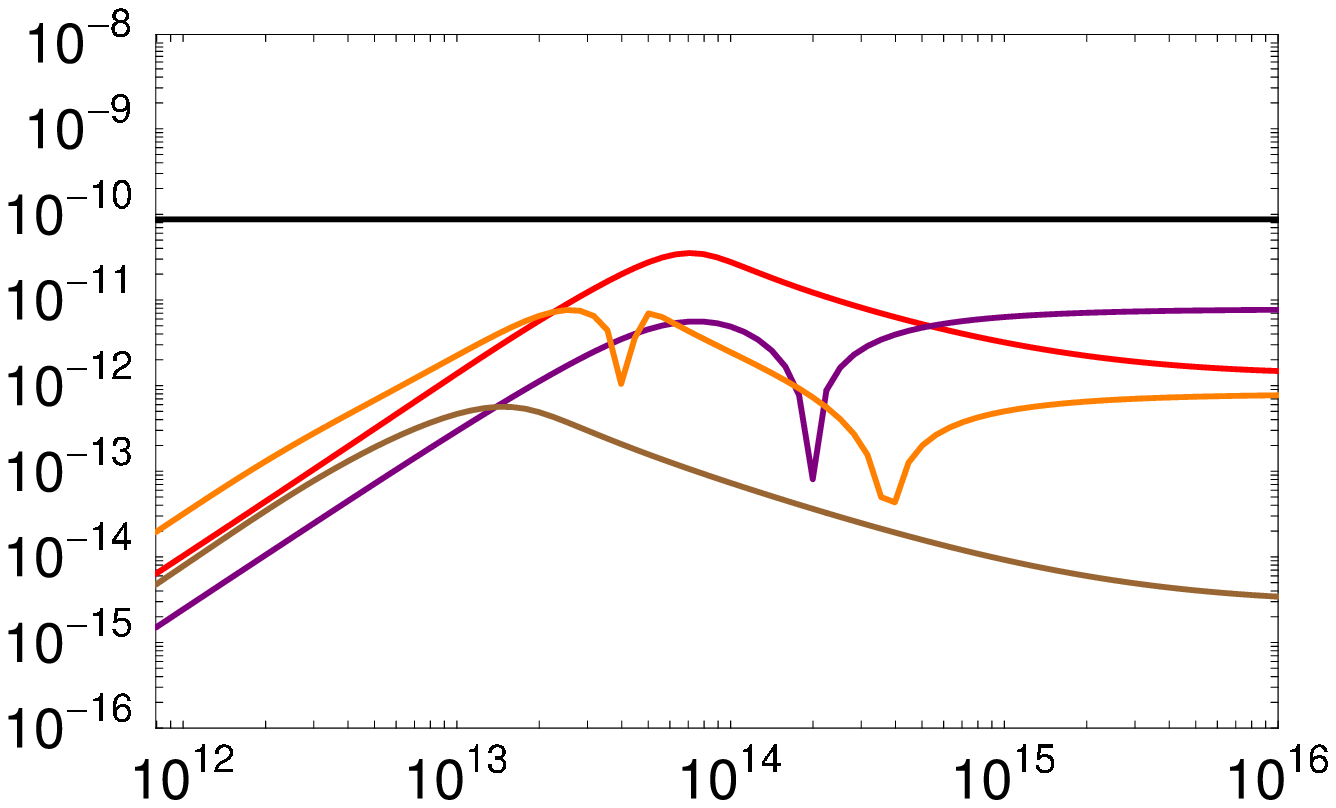}
\end{center}
\vskip -.1cm
\caption{Same as Fig.~\ref{fig:YB}, but with corrections to the relation
$M_d = M_e$ from the non-renormalizable operators
${\bf 16_i} {\bf 16_j} {\bf 10_d} {\bf 45}$, keeping
the relation $M_D = M_u$. Four
different choices of the matrix $U_m$ and of the CP-violating phases.}
\label{fig:YB_Um}       
\end{figure*}

However, this is not the whole story, since the above results
were obtained assuming the $SO(10)$ mass relation
$M_d = M_e$, which is in gross conflict with experimental data.
Corrections to this formula, e.g. from non-renormalizable
operators of the form ${\bf 16_i} {\bf 16_j} {\bf 10_d} {\bf 45}$,
will modify the reconstructed $f_{ij}$'s by introducting a
mismatch $U_m$ between the bases of charged lepton
and down quark mass eigenstates. Fig.~\ref{fig:YB_Um} shows
how the final baryon asymmetry is modified when the effect
of $U_m$ is taken into account. We can see that several
choices for $U_m$ (the measured charged lepton and down
quark masses do not fix all parameters in $U_m$)
lead to successful leptogenesis in the $(+,-,+)$ case, but not
in the $(-,-,-)$ case. There is some tension between
successful leptogenesis and gravitino overproduction in the
$(+,-,+)$ solution but, exactly as in the $(+,+,+)$ solution,
the observed asymmetry is generated over a significant portion
of the parameter space with $M_1 < 10^{10}$ GeV.


\section{Conclusions}
\label{sec:conclusions}

We have studied leptogenesis in supersymmetric $SO(10)$ models
with a left-right symmetric seesaw mechanism, including
flavour effects and the contribution of the next-to-lightest
right-handed neutrino. Assuming the relation $M_D = M_u$
and a hierarchical light neutrino mass spectrum, we found that
the ``type II-like'' solutions $(+,+,+)$ and $(-,+,+)$, as well as
the solutions $(+,-,+)$ and $(-,-,+)$, can lead to successful leptogenesis.
An accurate description of charged fermion masses was
a crucial ingredient in the analysis.
By contrast, the solution $(-,-,-)$ fails to generate the observed
baryon asymmetry from $N_2$ decays,
and a similar conclusion holds for the 3 other solutions with
$x_3 = x^-_3$ if one requires $M_1 < 10^{10}$ GeV.

Some comments about the generality of our results are in order:
(i) Although the above results were obtained for $M_D = M_u$, the same
  qualitative behaviour of the 8 solutions is expected for a more generic
  hierarchical Dirac matrix. Of course, whether leptogenesis is successful
  or not in a given solution can only be decided on a model-by-model basis;
(ii) At the quantitative level, different input parameters (other than
  the various phases and $U_m$) can significantly affect the results
  presented in Figs.~\ref{fig:spectra} to~\ref{fig:YB_Um}. This is most
  notably the case of the light neutrino mass parameters: $\theta_{13}$,
  $m_1$ and the type of the mass hierarchy (see Ref.~\cite{AHJL}
  for details). Also, corrections to the relation $M_D = M_u$ could
  have a significant impact, since e.g.  both $M_1$ in the $(+,-,+)$
  solution and $M_2$ in the $(-,-,-)$ solution are proportional
  to $y^2_2 v^2_u / m_3$.


\subsection*{Acknowledgements}

\vskip -.1cm

This work has been supported in part by the RTN European Program 
MRTN-CT-2004-503369, the Marie Curie Excellence Grant
MEXT-CT-2004-014297, and the French Program
``Jeunes Chercheurs'' of the Agence Nationale de la Recherche
(ANR-05-JCJC-0023).
PH and SL would like to thank Carlos Savoy for a pleasant and fruitful
collaboration on Ref.~\cite{HLS}.


%
%

\end{document}